\documentclass{ws-ijmpd} 

\begin{document}

\markboth{M.J. Rebou\c{c}as}  
{Constraining density parameters from cosmic topology}

\catchline{}{}{}{}{}

\title{CONSTRAINTS ON THE COSMOLOGICAL DENSITY \\
                  PARAMETERS AND COSMIC TOPOLOGY}

\author{M.J. REBOU\c{C}AS}

\address{Centro Brasileiro de Pesquisas F\'{\i}sicas  \\
Rua Dr.\ Xavier Sigaud 150 \\ 22290-180 Rio de Janeiro -- RJ, Brazil \\
reboucas@cbpf.br}


\maketitle

\begin{history}
\received{Day Month Year}
\end{history}

\begin{abstract}
A nontrivial topology of the spatial section of the universe is an 
observable, which can be probed for all homogeneous and isotropic 
universes, without any assumption on the cosmological density 
parameters. We discuss how one can use this observable to set 
constraints on the density parameters of the universe by using a 
specific spatial topology along with type Ia supenovae and X-ray 
gas mass fraction data sets. 
\end{abstract}


\section{Introduction}    

The standard Friedmann--Lema\^{\i}tre--Robertson--Walker (FLRW) 
approach to model the Universe commences with two basic assumptions. 
First, it is postulated the existence of a cosmic time $t$. Second, 
it is assumed that our $3$--dimensional space is homogeneous and 
isotropic. The most general spacetime metric consistent with 
these assumptions is
\begin{equation}
\label{RWmetric} ds^2 = -dt^2 + a^2 (t) \left [ d \chi^2 +
S_k^2(\chi) (d\theta^2 + \sin^2 \theta  d\phi^2) \right ] \;,
\end{equation}
where $S_k(\chi)=(\chi\,$, $\sin\chi$, $\sinh\chi)$ depending on
the sign of the constant spatial curvature ($k=0,1,-1$), and $a(t)$
is the cosmological scale factor.  The metric~(\ref{RWmetric}) only 
expresses the above assumptions, and to proceed further in this 
geometrical approach to model the physical world, one needs
a metrical theory of gravitation as, for example, General
Relativity (GR), which we assume in this article, to 
study the dynamics of the Universe.

However, GR is a metrical (local) theory, which relates the matter 
content of the Universe to its geometry. It does not specify the 
underlying  spacetime manifold $\mathcal{M}_4$ nor the corresponding 
spatial ($t=$constant) section $M$. This is the very first origin 
of the so-called cosmic topology in the context of the FLRW 
cosmological modelling, and arises from the simple fact that geometry 
constrains but does not dictate topology (see, e.g., the review 
articles Ref.~\refcite{CosmTopReviews}).
To illustrate this fact in a very simple way, imagine 
a two-dimensional ($2$--D) world and its beings. Suppose further that 
these $2$--D creatures have a ($1+2$)--spacetime geometrical theory 
of gravitation, and that modelling their universe in the framework of 
this theory they found that they live in a flat universe, i.e. that the 
$2$--D spatial geometry is Euclidean. 
This knowledge, however, does not give them enough information to 
determine the space topology of their world. Indeed, besides the 
simply-connected Euclidean plane $\mathbb{R}^2$, the space section of 
their universe can take either of the following multiply-connected 
space forms: the cylinder  
$\mathbb{C}^2 = \mathbb{R} \times \mathbb{S}^1$, the torus 
$\mathbb{T}^2= \mathbb{S}^1 \times \mathbb{S}^1$, the Klein bottle 
$\mathbb{K}^2 = \mathbb{S}^1 \times \mathbb{S}^1_{\pi}$ and the 
M\"obius band $\mathbb{M}^2 = \mathbb{R} \times \mathbb{S}^1_{\pi}$. 

Similarly in the  FLRW  approach to model the our ($1+3$)--dimensional 
world, although the spatial section $M$ is usually taken to be one of 
the simply connected spaces: Euclidean $\mathbb{E}^3$, spherical 
$\mathbb{S}^3$ or hyperbolic $\mathbb{H}^3$, it is a mathematical 
result that the great majority of locally homogeneous and isotropic 
$3$--spaces $M$ are multiply-connected manifolds, i.e. spaces with a 
nontrivial topology. 
Thus, for example, in a universe whose geometry of the spatial section is 
Euclidean ($k=0$), besides $\mathbb{E}^{3}$ there are 10 classes 
of topologically distinct compact $3$--spaces $M$ that admits this geometry,
while for universes with either spherical ($k=1$) and hyperbolic
($k=-1$) spatial geometries there is an infinite number of topologically 
inequivalent (non-homeomorphic) manifolds with non-trivial topology that 
can be endowed with these geometries.  

In a FLRW models the sign of the spatial curvature $k$, and therefore 
the associated geometry, is given by the total matter-energy density 
$\Omega_0$ of the Universe, through the Friedmann equation 
in the form $k= H_0^2 a_0^2 (\Omega_0-1)$.%
\footnote{For details on the notation see section~\ref{PrereqSett}.}
Indeed, for $\Omega_0 > 1$ the spatial section is positively 
curved (spherical geometry), for $\Omega_0 = 1$ it is flat ($k=0$), 
while for $\Omega_0 < 1$ the section $M$ is negatively curved
(hyperbolic geometry).
As a consequence, a key point in the search for the (spatial) geometry 
of the Universe is to constrain the total density  $\Omega_0$ from
observations. In the context of the $\Lambda$CDM  model, which we
assume in this article, this amounts to determining regions in the 
$\Omega_{m}\,$--$\,\,\Omega_{\Lambda}$ parametric plane, which 
consistently account for the observations, and from which one expects 
to deduce the geometry of the Universe.

Now given that the spatial geometry is an observable that constrains 
but does not determine the topology of the $3$--space $M$, two
questions arise at this point, namely whether the topology is an 
observable, and, if so,  to what extent it can be used to either 
determine the geometry or set constraints on the density parameters 
$\Omega_m$ and $\Omega_{\Lambda}$. 

Our main aims here, which are based upon and complementary to
our previous works,\cite{PreviousRAMM}\cdash\cite{PreviousRAb} 
(see also Refs.~\refcite{PreviousMMR,PriviousBBRS,PriviousBBRSantos}) 
are twofold: first, we point 
out that a nontrivial topology of the spatial sections is as an
observable attribute as local curvature, and can be probed for 
all homogeneous and isotropic universes with no assumption on the 
cosmological density parameters; second, we demonstrate that the 
knowledge of the spatial topology allows to place constraints on 
the density parameters of the Universe.

The structure of the paper is as follows. In the next section,
to make the article as clear and self-contained as possible,
we review a few topological results regarding $3$-manifolds, 
and give an account of the cosmological model employed. 
In section~\ref{MainRes}, we discuss how one can use an observable 
nontrivial topology of the spatial section of the Universe 
to set constraints on the cosmological density parameters, and
present explicit examples of such bounds. 

\section{Topological Prerequisites and Cosmological Setting}
\label{PrereqSett}

Within the framework of the standard FLRW cosmology, the Universe is
modelled by a $4$-manifold $\mathcal{M}_4$ which is decomposed into
$\mathcal{M}_4 = \mathbb{R} \times M$, and is endowed with a locally
isotropic and homogeneous Robertson--Walker (RW) metric~(\ref{RWmetric}).
The spatial section $M$ is usually taken to be one of the following
simply-connected spaces: Euclidean $\mathbb{E}^{3}$ ($k=0$),
spherical $\mathbb{S}^{3}$ ($k=1$), or hyperbolic $\mathbb{H}^{3}$
($k=-1$) spaces. However, since geometry does not dictate topology,
the $3$-space $M$ may equally well be any one of the possible
quotient (multiply-connected) manifolds $\mathbb{E}^3/\Gamma$, 
$\mathbb{S}^3/\Gamma$, and $\mathbb{H}^3/\Gamma$, where $\Gamma$ is 
a fixed-point free discrete group of isometries of the corresponding 
covering space $\mathbb{E}^3$, $\mathbb{S}^3$, or $\mathbb{H}^3$.

Quotient manifolds are compact in three independent directions with 
no boundary (referred to, by mathematicians, as closed), or 
compact in two or at least one independent direction.
The action of $\Gamma$ tiles the corresponding covering space  
$\mathbb{E}^3$, $\mathbb{S}^3$, or $\mathbb{H}^3$, into 
identical cells or domains which are copies of the so-called fundamental 
polyhedron (FP). A FP plus the face identifications given by the group 
$\Gamma$ is a faithful representation of the quotient manifold $M$.
An example of quotient manifold in three dimensions is the flat $3$--torus 
$T^3=\mathbb{S}^1 \times \mathbb{S}^1 \times \mathbb{S}^1=\mathbb{E}^3/\Gamma$.
The  covering space clearly is $\mathbb{E}^3$, and a FP is a cube with 
opposite faces identified after a translation. This FP tiles the covering 
space $\mathbb{E}^3$. The group $\Gamma=\mathbb{Z} \times \mathbb{Z} \times 
\mathbb{Z}$ consists of discrete translations associated with the face 
identifications.

In a multiply-connected manifold, any two given points may be joined
by more than one geodesic. Since the radiation emitted by cosmic
sources follows geodesics, the immediate observational consequence 
of a detectable non-trivial spatial topology of $M$ is that the sky 
will show multiple images of either cosmic objects or 
specific spots of the cosmic microwave background radiation (CMBR). 
At very large scales, the existence of these multiple images 
(or pattern repetitions) is a physical effect that can be used to 
probe the $3$-space topology. We shall return to this point in the
next section.

An important topological length of the spherical and hyperbolic 
$3$--manifolds $M$ is the so-called injectivity radius $r_{inj}$, 
which is nothing but the radius of the smallest sphere `inscribable' 
in $M$, and can be formally defined in terms of the length of the 
smallest closed geodesics $\ell_M\,$ by $r_{inj} = \ell_M/2$ (for 
details on the formal definition of $r_{inj}\,$ see, e.g., 
Ref.~\refcite{rinj}).

Motivated by the best fit value of the total density $\Omega_0
=1.02 \pm\, 0.02$ ($1\sigma$ level) reported by WMAP team,\cite{WMAP-Spergel} 
which includes a positively curved universe as a realistic possibility,
in this work we focus our attention in globally homogeneous spherical 
manifolds, of which we shall recall some relevant results in what
follows. 

The multiply connected spherical $3$-manifolds are of the form 
$M=\mathbb{S}^3/\Gamma$, where $\Gamma$ is a finite fixed-point 
free subgroup of $SO(4)$. These manifolds were originally classified 
in Ref.~\refcite{ThrelfallSeifert} (for a description in the context 
of cosmic topology see the pioneering article by Ellis\cite{Ellis71}). 
Such a classification consists essentially in the enumeration of all finite 
groups  $\Gamma \subset SO(4)$, and then in grouping the possible manifolds 
in classes. In a recent paper,\cite{GLLUW} the classification has been 
recast in terms of single action, double action, and linked action 
manifolds. Single action manifolds are globally homogeneous, and then 
satisfy a topological principle of (global) homogeneity, in the sense 
that all points in $M$ share the same topological properties. 
In Table~\ref{SingleAction} we list the single action manifolds 
together with the symbol we use to refer to them, the covering groups
$\Gamma$ and their order as well as the corresponding injectivity radius
$r_{inj}$. 
Finally we note that the binary icosahedral group $I^{\ast}$ gives rise 
to the known Poincar\'e dodecahedral space $\mathcal{D}$, whose fundamental 
polyhedron (FP) is a regular spherical dodecahedron, $120$ of which tile the 
$3$-sphere into identical cells which are copies of the FP.

\begin{table}[ph]
\tbl{Single action spherical manifolds together with their names, 
the covering groups and their order, and the injectivity radius
$r_{inj}$.}
{\begin{tabular}{@{}cccc@{}} \toprule
 Manifold  & Covering Group $\Gamma$ & Order of $\Gamma$ & Injectivity Radius \\
\colrule
$\mathcal{Z}_n:=\mathbb{S}^3/Z_n$    &   Cyclic            $Z_n$   & $n$  & $\pi/n$  \\
$\mathcal{D}^*_m:=\mathbb{S}^3/D^*_m$& Binary dihedral     $D^*_m$ & $4m$ & $\pi / 2m$ \\
$\mathcal{T}:=\mathbb{S}^3/T^*$      & Binary tetrahedral  $T^*$   & 24   & $ \pi/6$  \\
$\mathcal{O}:=\mathbb{S}^3/O^*$      & Binary octahedral   $O^*$   & 48   & $\pi/8$  \\
$\mathcal{D}:=\mathbb{S}^3/I^*$      & Binary icosahedral  $I^*$   & 120   & $\pi/10$  \\
\botrule
\end{tabular} \label{SingleAction} }
\end{table}

An important point regarding the spherical manifolds is that the injectivity 
radius $r_{inj}$ expressed in \emph{units of the curvature radius} (terms that
we define below) is a constant (topological invariant) for a given manifold 
$M$.

Let us examine now our cosmological assumptions regarding the cosmic
constituents, and discuss the chief point in the search for the 
(spatial) geometry of the Universe. In the light of current observations, 
we assume that the current matter content of the Universe is well 
approximated by cold dark matter (CDM) of density $\rho_m$ plus a 
cosmological constant $\Lambda$. The Friedmann equation is then 
given by
\begin{equation}
\label{Feq}
H^2 =\frac{8 \pi G \rho_m }{3} -\frac{k}{a^2}
            +\frac{\Lambda}{3}\;,
\end{equation}
where $H=\dot{a}/a$ is the Hubble parameter and $G$ is
Newton's constant. Introducing
\begin{equation}
\rho^{}_{crit} := \frac{3 H^2}{8\pi G}\;, \qquad
\Omega_m := \frac{\rho_m}{ \rho^{}_{crit}} \;, \qquad
\Omega_{\Lambda}: = \frac{\rho_{\Lambda}}{ \rho^{}_{crit}}=
                                \frac{\Lambda}{3 H^2}\,\,,
\end{equation}
equation~(\ref{Feq}) gives
\begin{equation}
\label{Friedmann}
k= H^2 a^2 (\Omega_m + \Omega_{\Lambda}- 1)= 
 H^2 a^2 (\Omega_0-1) \;,
\end{equation}
where clearly $\,\Omega_0 = \Omega_m + \Omega_{\Lambda}$.
As we have mentioned in the Introduction, Eq.~(\ref{Friedmann}) makes 
transparent that the spatial geometry of the Universe depend on whether 
the density parameters determine points $P= (\Omega_m,\, \Omega_{\Lambda})$
that lie in the regions below (hyperbolic geometry), above (spherical geometry), 
or on the flat line $\Omega_0 = \Omega_{\Lambda} + \Omega_{m}=1$ (Euclidean 
geometry).   

To conclude this section, we note that for non-flat metrics of the 
form~(\ref{RWmetric}), the scale factor $a(t)$ can be identified with 
the curvature radius of the spatial section of the Universe at time 
$t=t_0$, which from~(\ref{Friedmann}) is clearly  given by 
$a_0 = |k|\,{H_0^{-1}\, |\Omega_0-1|^{-1/2}}$, where 
the subscript $0$ denotes evaluation at present time $t_0$. 
Thus, for non-euclidian spatial geometries the distance $\chi$ of any 
point with coordinates $(\chi, \theta, \phi)$ to the origin in 
\emph{units of the curvature radius}, $a_0=a(t_0)$, is given by 
\begin{equation}
\label{redshift-dist}
\chi = \frac{d}{a_0}=\sqrt{|1- \Omega_0|} \int_0^{z} 
\frac{dx}{\sqrt{(1+x)^3 \Omega_{m0} + (1+x)^2 (1- \Omega_0) + 
 \Omega_{\Lambda 0}}} \;\, ,
\end{equation}
In this paper we measure the lengths in unit of curvature 
radius $a_0$.

\section{Main Results and Concluding Remarks}
\label{MainRes}

Given that the spatial geometry is an observable that constrains but does 
not dictate the topology of the $3$--space $M$, two questions arise 
at this point: whether the topology is an observable, and, if so, 
whether it can be used to set constraints on the density parameters 
associated with dark matter ($\Omega_m$) and dark energy ($\Omega_{\Lambda}$).
Regarding the former, over the past few years many strategies and methods 
to probe a nontrivial spatial of the $3$--dimensional space, by using 
either discrete cosmic sources or CMBR, have been devised (besides the 
review articles Refs.~\refcite{CosmTopReviews} see also, e.g., Refs.%
~\refcite{TopSign} and references therein). 

To make clear that a nontrivial spatial topology is as an observable
attribute as the local curvature, we shall focus on the so-called 
``circles-in-the-sky" method, which relies on multiple 
copies of correlated circles in the CMBR maps,\cite{CSS1998} and
can be briefly described as follows.
In a space with a detectable non-trivial topology,%
\footnote{A detailed discussion on the detectability of cosmic
topology can be found in Refs.~\refcite{rinj} and~\refcite{TopDetec}.}
the last scattering sphere (LSS) intersects some of its topological images 
along pairs of circles of equal radii, centered at different points on 
LSS, with the same distribution of temperature fluctuations, $\delta T$. 
Since the mapping from the LSS to the night sky sphere preserves 
circles,\cite{CGMR05} these pairs of matching circles will be inprinted 
on the CMBR temperature fluctuations sky maps regardless of the background 
geometry and detectable topology.
As a consequence, to observationally probe a non-trivial topology 
one should scrutinize the full-sky CMBR anisotropy maps in order to extract 
the correlated circles, whose angular radii and relative position 
of centers can be used to determine the spatial topology of the 
Universe. Thus, a non-trivial topology of the space section of the 
Universe is an observable, and can be probed, through the 
circles-in-the-sky, for all locally homogeneous and isotropic 
universes with no assumption on the cosmological density parameters.

Regarding the question as to whether this observable can be used to
set constraints on the density parameters, we first recall that it 
is known that the topology of a constant curvature $3$--dimensional 
manifold determines the sign of its curvature (see, e.g., 
Ref.~\refcite{BernshteinShvartsman1980}).
Thus, the topology of the spatial section $M$ dictates its 
geometry. 
At first sight this seems to indicate  that the detection of
the spatial topology gives rise to very loose constraints
on both density parameters, since it only determines whether they
take values in the regions below, above, or on the flat line $\Omega_0=\Omega_{m}+\Omega_{\Lambda}=1$. 
However, in what follows we show that the knowledge of spatial 
topology through the ``circles-in-the-sky" gives rise to very 
remarkable constraints on the density parameters allowed by 
other observational data sets.

Motivated by the best fit value of the total density $\Omega_0 =1.02 
\pm\, 0.02$ ($1\sigma$ level) reported by WMAP team,\cite{WMAP-Spergel} 
which includes a positively curved universe as a realistic possibility, 
we consider here examples of globally homogeneous spherical manifolds,
which we have discussed in the previous section. One of these 
topologically homogeneous spaces is the Poincar\'e dodecahedral space 
$\mathcal{D}$, which has been suggested by Luminet \emph{et al.}\cite{Poincare} 
as a possible explanation for the observed anomalous power of the low 
multipoles,\cite{WMAP-Spergel} and since then it has been the scope of further investigations\cite{Cornish}\cdash\cite{Aurich3} (
for alternative studies of the missing 
wide-angle correlations see, e.g., Ref.~\refcite{Wide-Angle_Refs} ).

In Ref.~\refcite{Aurich2} a systematic study of the CMBR anysotropy
in the spherical globally homogeneous spaces has been made. They have 
found that $\mathcal{D}$, $\mathcal{O}$ and $\mathcal{T}$ out of 
infinitely many manifolds account for the suppression of power at 
large scales observed by WMAP,\cite{WMAP-Spergel} and also fit the WMAP  
temperature two-point correlation function. But, for the total density 
$\Omega_{0}$ restricted to the WMAP interval (1.00, 1.04], 
the space $\mathcal{T}$ is excluded since it requires a value of 
$\Omega_{0}$ outside this interval. On the other hand, more recently 
they have carried out a detailed search for the circles-in-the-sky 
in these three manifolds,\cite{Aurich3} but due to noise and foreground 
composition, only a non-conclusive indication for the correlated circles 
has been reported for the spaces $\mathcal{D}$ and $\mathcal{T}$.
Here we restrict our examples to the FLRW models with spatial section 
$\mathcal{T}$. 

\begin{figure}[!htb]
\centerline{\def\epsfsize#1#2{0.45#1}\epsffile{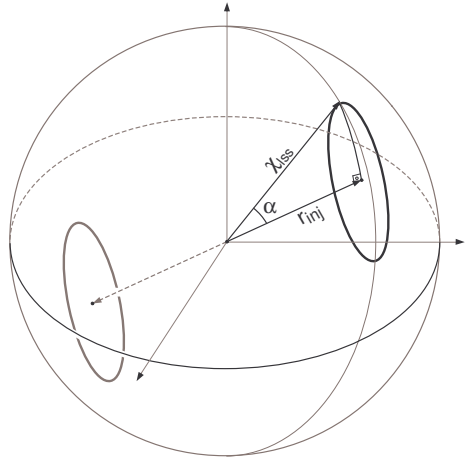}}
\caption{A schematic illustration of two antipodal
matching circles in the sphere of last scattering. These pairs of circles 
occur in all globally homogeneous positively curved manifolds with a 
detectable nontrivial topology. The relation between the angular radius 
$\alpha$ and the angular sides $r_{inj}$ and $\chi^{}_{lss}$ is given 
$\cos \alpha = \tan r_{inj}\, \cot \chi^{}_{lss}$.\label{CinTheSky1}}
\end{figure}

In globally homogeneous spherical $3$--spaces the matching circles 
have to be antipodal, i.e. the centers of correlated circles are separated 
by $180^\circ$, as shown in Fig.~\ref{CinTheSky1}.
Clearly the distance between the centers of each pair of the {\it first}
correlated circles is twice the injectivity radius $r_{inj}$.
Now, a straightforward use of trigonometric relations
(known as Napier's rules) for the right-angled 
spherical triangle shown in Fig.~\ref{CinTheSky1} yields 
\begin{equation}
\label{chilss}
\chi^{}_{lss} = \tan^{-1} \left[\,\frac{\tan r_{inj}}{
\cos \alpha}\, \right] \;,
\end{equation}
where $r_{inj}$ is a topological invariant, whose values are given
in Table~\ref{SingleAction}, and the distance $\chi^{}_{lss}$ of
the last scattering surface to the origin in units of the curvature
radius is given by~(\ref{redshift-dist}) with $z=z_{lss}=1089$.%
\cite{WMAP-Spergel}

Equation~(\ref{chilss}) along with (\ref{redshift-dist}) give 
the relation between the angular radius $\alpha$ and the cosmological 
density parameters $\Omega_{\Lambda}$ and $\Omega_{m}$, and thus can be
used to set constraints on these parameters. To quantify this we
proceed in the following way. Firstly, as an example, we assume the 
angular radius $\alpha = 50^\circ$.
Secondly, since the measurements of the radius $\alpha$
unavoidably involve observational uncertainties, in order
to obtain very conservative results we take $\delta {\alpha} \simeq
6^\circ$.%
\footnote{These are typical values for $\alpha$ and $\delta \alpha$.
However, our general conclusions hold regardless of the precise value 
for $\alpha$ and its uncertainty.}

To illustrate the role of the cosmic topology in constraining the density 
parameters $\Omega_{m}\,$ and $\,\,\Omega_{\Lambda}\,$, we consider the 
binary tetrahedral $\mathcal{T}$ as the spatial topology to reanalyze 
(with this topological prior) the constraints on the density parameters
taking into account two data sets. First, the so-called \emph{gold} 
sample of 157 SNe Ia, as compiled by Riess \emph{et al.}\cite{Riess2004}
Second, the \emph{gold} sample along with the Chandra measurements of the 
X-ray gas mass fraction in 26 X-ray luminous, dynamically relaxed 
galaxy clusters provided by Allen {\it et al.}\cite{Allen2004}
The  $\mathcal{T}$ spatial topology is added to the conventional 
data analysis as a Gaussian prior on the value of $\chi^{}_{lss}$, 
which can be easily obtained from a elementary combination 
of~(\ref{chilss}) along with~(\ref{redshift-dist}).

\begin{figure}[!htb]
\centerline{\def\epsfsize#1#2{0.45#1}\epsffile{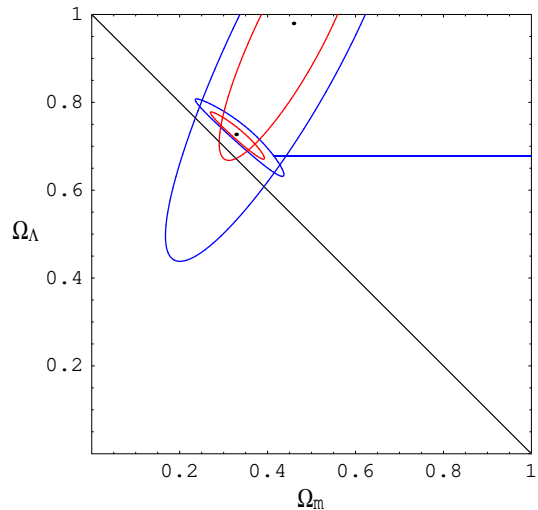}}
\caption{The 68.3\% and 95.4\% confidence regions in the density parametric 
plane, which arise from the SNe Ia plus binary tetrahedral space topology 
analysis. The best fit values for the density parameters are 
$\Omega_m =  0.33^{+0.09}_{-0.08}$, $\,\Omega_{\Lambda} = 0.73^{+0.07}_{-0.08}$ 
and $\Omega_0 = 1.05^{+0.03}_{-0.02} $ at 95.4\% confidence level. The 
conventional SNe Ia is also shown here for comparison.} 
\label{Top+SNIa}
\end{figure}

Figure~\ref{Top+SNIa} shows the results of our joint SNe Ia plus cosmic
topology analysis. There we display the confidence regions (68.3\%,
and 95.4\% confidence level, c.l.) in the parametric plane 
$\Omega_{m}\,$--$\,\,\Omega_{\Lambda}$ and also the regions from 
the conventional analysis.
The comparison between these regions makes clear that
the effect of the $\mathcal{T}$ topology as a new cosmological
observable is to reduce considerably the area corresponding to the
confidence intervals in the parametric plane as well as to break 
degeneracies arising from the current SNe Ia measurements.
For a detail study of the role play by the globally homogeneous 
topologies $\mathcal{D}$, $\mathcal{O}$ and $\mathcal{T}$, we
refer the Ref.~\cite{AMMR2006}.
 
\begin{figure*}[t]
\centerline{
\psfig{figure=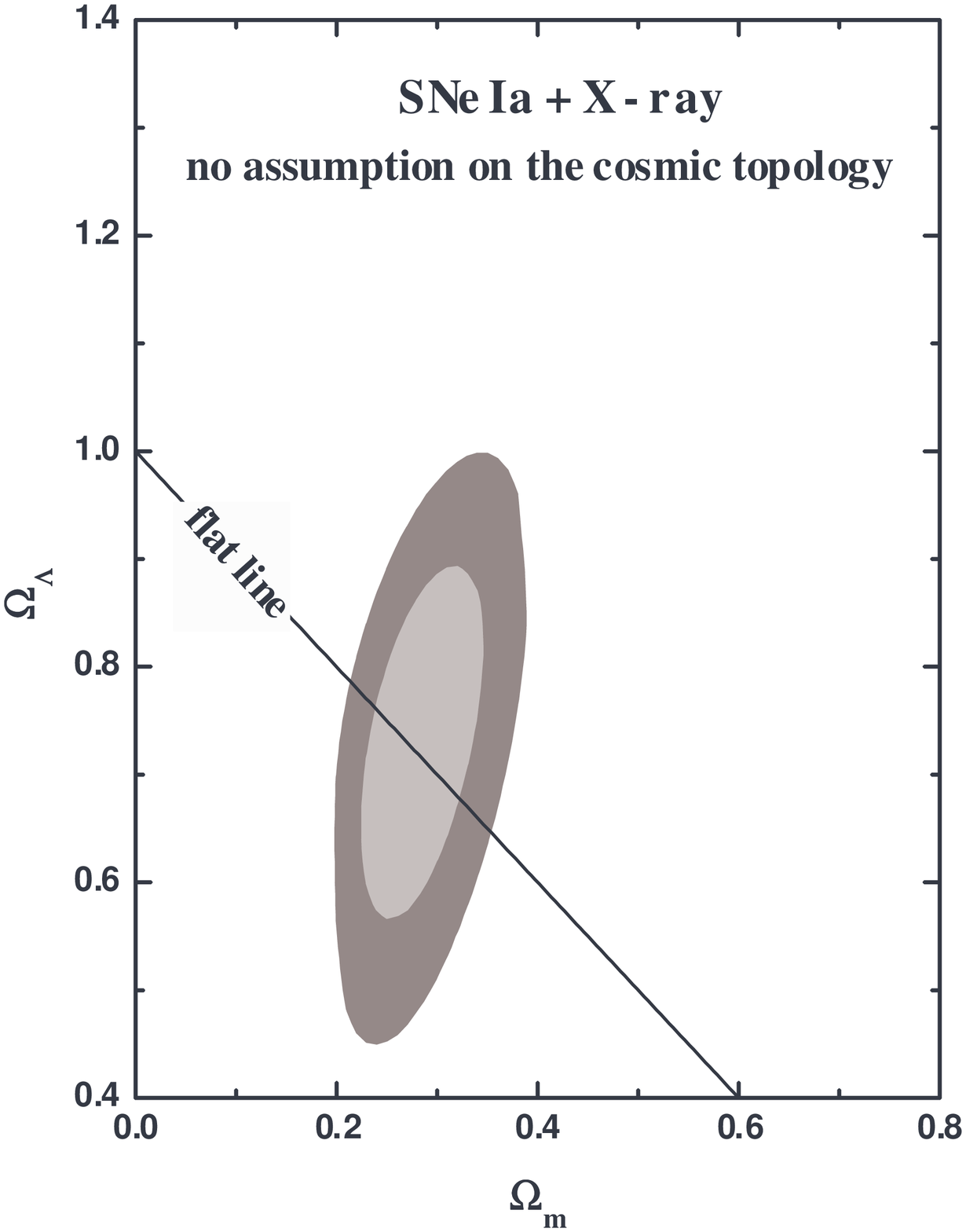,width=2.4truein,height=2.8truein,angle=0} \hskip -0.2cm
\psfig{figure=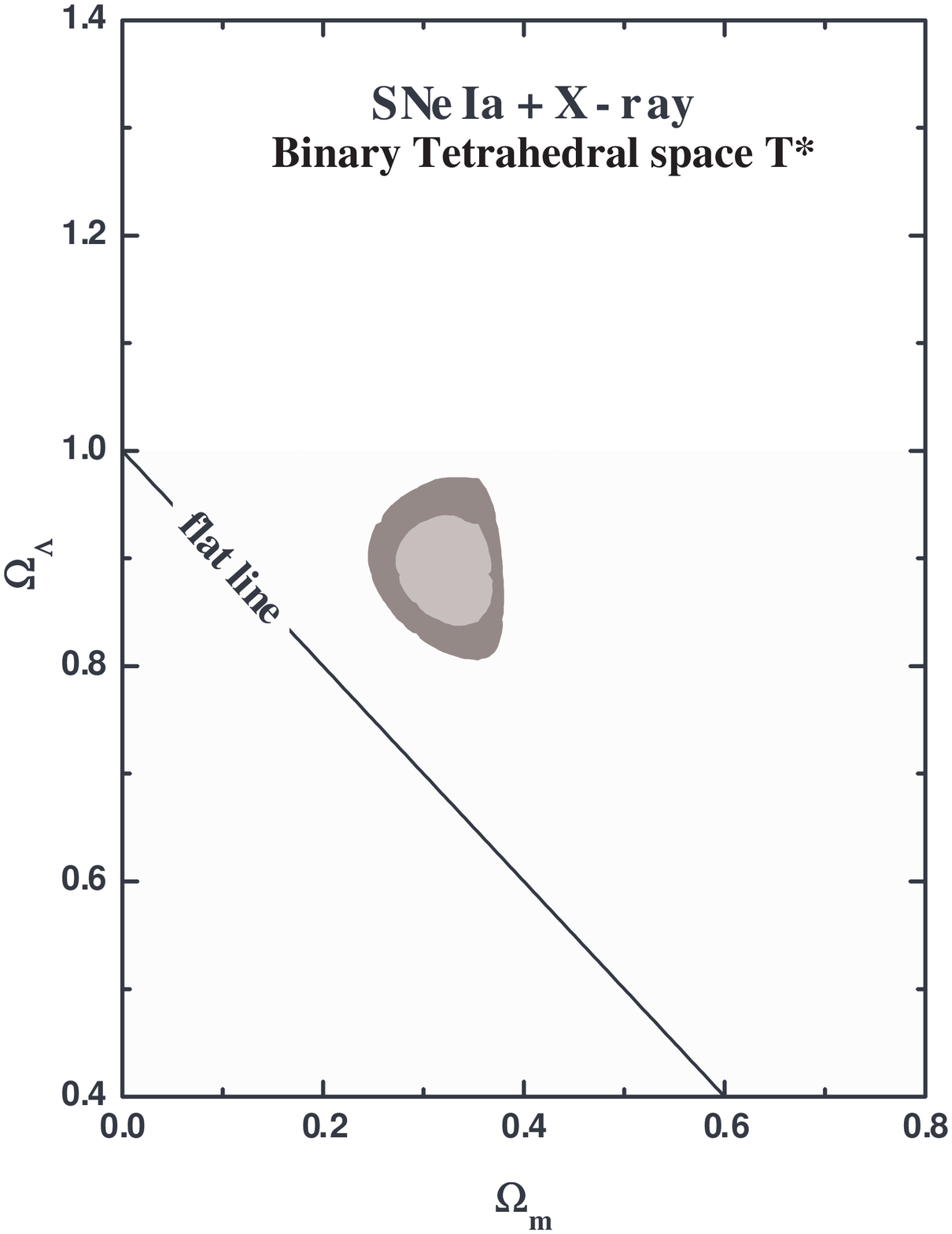,width=2.4truein,height=2.8truein,angle=0}\vspace{-0.1cm}}
\caption{The results of our statistical analyses. The panels show confidence 
regions ($68.3\%$ and $95.4\%$ c.l.) in the $\Omega_m - \,\,\Omega_{\Lambda}$ 
plane from the latest Chandra measurements of the X-ray gas mass fraction in 
26 galaxy clusters ($0.07 < z < 0.9$) plus determinations of the baryon density 
parameter, measurements of the Hubble parameter and the \emph{gold} sample of 
157 SNe Ia. The left panel shows the results of the conventional SNe Ia plus 
X-ray analysis, while the right panel contains the result of SNe Ia+X-ray+topology
for the spatial topology $\mathcal{T}$. For this latter analysis 
at 95.4\% c.l., the best fit values for the density parameters are 
$\Omega_m = 0.32 \pm 0.06$, $\,\Omega_{\Lambda} = 0.89 \pm 0.06$,
and $\Omega_0 = 1.21 \pm 0.08$.} 
\label{StatAnalysis}
\end{figure*}

Figure~\ref{StatAnalysis} shows the results of our joint SNe Ia+X-ray plus 
topology analysis. Confidence regions (68.3\% and 95.4\% c.l.) 
in the parametric space $\Omega_{m}\,$--$\,\,\Omega_{\Lambda}\,$ are 
displayed.
For the sake of comparison, we also show in the left panel the density 
parametric plane for the conventional SNe Ia plus Galaxy Clusters 
analysis (without the cosmic topology assumption). By comparing 
these analyses, it is clear that again for these data sets the 
$\mathcal{T}$ non-trivial space topology reduces considerably the 
parametric space region allowed by the current observational data, 
and breaks some degeneracies arising from the current SNe Ia and X-ray 
gas mass fraction measurements. 
The best-fit parameters for this SNe Ia+X-ray+topology analysis provides 
$\Omega_m = 0.32 \pm 0.06$,  $\,\Omega_{\Lambda} = 0.89 \pm 0.06$ and
$\Omega_0 = 1.21 \pm 0.08$.

Concerning the above analyses it is worth emphasizing  three important 
features. First, that the best-fit values are just weakly dependent on 
the value used for the angular radius $\alpha$ of the circle.
Second, the uncertainty $\delta \alpha$ alters predominantly the area 
of the confidence regions, without having a significant effect on the 
best-fit values. 
Third, there is a topological degeneracy in that the same best fits 
and  confidence regions for the $\mathcal{T}$ topology, would 
arise from either $\mathcal{Z}_{6}$ or $\mathcal{D}^*_3$ topology.

\section*{Acknowledgments}
I thank CNPq for the grant under which this work was carried out.
I am grateful to J.S. Alcaniz,  M. Makler and B. Mota for their
valuable help with the figures.
I also thank the organizers of IWARA 2005, in particular to 
C\'esar Z. Vasconcellos and Bardo E.J. Bodmann for the invitation, 
hospitality, and for the friendly and successful atmosphere of the 
workshop. 
I am grateful to A.F.F. Teixeira for reading the manuscript and 
indicating misprints and omissions.

\end{document}